\documentclass[tightenlines,aps,nofootinbib]{revtex4}

\usepackage{psfig,axodraw,float}
\usepackage{graphics}

\def\as{\alpha_s}

\def\Q{{\mathcal{Q}}}
\def\D{{ D}^{\rm ini}}
\def\MSbar{\overline{\mathrm{MS}}}
\def\Se{S_\epsilon}
\def\e{\epsilon}

\def\Li{{\rm Li}}

\begin{document}

\begin{flushright}
\vbox{
\begin{tabular}{l}
UH-511-1048-04\\
\   \end{tabular} }
\end{flushright}

\vspace{0.6cm}

\title{
Perturbative Heavy Quark Fragmentation Function
through ${\cal O}(\alpha_s^2)$
}

\author{Kirill Melnikov $\!\!$\thanks{
e-mail:  kirill@phys.hawaii.edu}
and  Alexander Mitov
}
\affiliation{Department of Physics and Astronomy,\\ University of Hawaii,\\
Honolulu, HI, 96822}

\begin{abstract}

\vspace{2mm}

We derive the initial condition for the perturbative fragmentation
function of a heavy quark through order ${\mathcal{O}}(\as^2)$ in
the $\overline {\rm MS}$ scheme. This initial condition  is useful
for computing heavy quark (or lepton, in case of QED) energy
distributions from calculations with massless partons. In
addition, the initial condition at ${\cal O}(\alpha_s^2)$ can be
used to resum collinear logarithms $\ln(Q^2/m^2)$ in heavy quark
energy spectrum with next-to-next-to-leading logarithmic accuracy
by solving the DGLAP equation.
\end{abstract}

\maketitle

\thispagestyle{empty}

\section{Introduction}
Production  of heavy flavors (charm and bottom) in high energy
processes has become an  important subject in the last decade, due
to experiments at $e^+e^-$  and hadron colliders. Since the center
of mass energy of these machines is much larger than bottom and
charm quark masses, it is tempting to consider such processes in
the massless approximation. Unfortunately, this is only possible
for sufficiently inclusive  observables, for example, total
cross-sections. However, differential distributions are  more
relevant for experimental analysis and, in addition, contain more
information about the underlying physics. An interesting example
is the energy distribution of heavy hadrons  produced in a
high-energy collision,  since it gives direct access to the
hadronization process. In perturbation theory, however, the heavy
quark energy distribution diverges in the limit $m\to 0$ and the
sensitivity to the quark mass $m$ remains at arbitrary high
energies.

This sensitivity is disagreeable for two reasons. First, it
implies that energy spectra must be computed  retaining quark
masses which  makes the  calculations quite
involved. Second, in higher orders of perturbation theory,
energy spectra  contain  powers of logarithms $\ln(Q^2/m^2)$, that
may  invalidate  fixed order perturbative calculations.

Both of these problems are solved by introducing perturbative
fragmentation function \cite{MN}. This function describes the
probability that a massless parton of a certain type fragments
into a heavy quark.  The mass of the heavy quark  provides a
natural cut-off for the collinear radiation, thus making
perturbative predictions finite. The perturbative fragmentation
function satisfies the  DGLAP evolution equation \cite{DGLAP}. By
solving this equation, we can sum up large collinear logarithms
$\ln(Q^2/m^2)$ that appear in perturbative fragmentation function
and  improve the quality of perturbative calculations. To avoid
confusion, we note that a meaningful prediction for heavy hadron
energy spectrum  can only be obtained if the  heavy quark energy
distribution is convoluted with a non-perturbative fragmentation
function. Such functions are traditionally extracted by fitting
hadron energy spectra in $e^+e^-$ collisions.  We will not discuss
this issue here;  for recent work on the subject see
\cite{PFFapplications}. We also  note that the concept of
perturbative fragmentation is useful outside QCD; for example, QED
effects in  the electron energy spectrum in muon decay, currently
under study by the TWIST collaboration \cite{twist}, can be
analyzed using the same technique \cite{technique}.

Because the $Q^2$ evolution of the fragmentation function can be
obtained from the DGLAP equation, the perturbative fragmentation
function at arbitrary $Q^2$ can be fully reconstructed once the
initial condition  $\D$  and the time-like splitting kernels for
the DGLAP evolution are known. The initial condition is usually
computed at $Q^2 \sim m^2$; this ensures that no large logarithms
appear and fixed order perturbation theory is reliable.

The purpose of this paper is to provide such initial condition for
the perturbative fragmentation function through ${\cal
O}(\alpha_s^2)$. As we explained earlier, the knowledge of $\D$
through next-to-next-to-leading order (NNLO) enables  the
calculation of energy  spectra of heavy quarks  through ${\cal
O}(\as^2)$ from massless results and, simultaneously, allows to
resum large logarithms $\ln(Q^2/m^2)$ using the DGLAP equation.
All power corrections ${\cal O}(m^2/Q^2)$ are neglected within
this approach, but for many practical purposes such precision is
sufficient.

Technically, the use of perturbative fragmentation functions makes
it possible to compute  the heavy quark energy spectrum in
two steps  -- we first compute the energy spectrum of
massless partons produced in a hard scattering and then convolute
this spectrum with perturbative fragmentation function.  The
possibility to remove all the dependence on the heavy quark mass
from the calculation of the hard scattering simplifies the
computations considerably.

The remainder of the paper is organized as follows. In Section II
we describe the process-independent derivation of the initial
condition for the fragmentation function. Then, we discuss
collinear factorization  when finite quark  masses are present
and explain how to compute $\D$ from relevant  Feynman diagrams.
We also suggest a suitable modification of the original proposal
\cite{KL,CC} for the process-independent computation of the
initial condition; such modification significantly simplifies NNLO
calculations. In Section IV the initial condition for the
fragmentation function  for both quark- and gluon-initiated
processes  is computed through NLO.
This computation allows us to demonstrate the details of
our approach and to  derive the  NLO perturbative fragmentation
function through $O(\epsilon)$, where $\epsilon$ is the
dimensional regularization parameter.
 In Section V we describe the calculation of the
${\cal O}(\alpha_s^2)$ contribution to the fragmentation function
and present the result for the initial condition.
We conclude in Section VI.

\section{Process-independent derivation of $\D$}

Consider production of a heavy quark $\Q$ with mass $m$ and a
definite value of energy $E_\Q$ in a hard scattering process.
According to the QCD factorization theorems
\cite{ftheorem,Ellis,purplebook}, the heavy quark energy spectrum
can be computed  as a convolution of the energy distribution of
massless partons produced in the hard process, and the
fragmentation function that describes the probability that the
massless parton fragments into a massive quark with a definite
energy. If the energy fraction $E_\Q/E_{\Q , max}$ of the heavy
quark is denoted by $z$, then the energy distribution of that
quark can be written as:
\begin{equation}
{d\sigma_\Q \over dz}(z,Q,m) = \sum_a\int_z^1{dx\over x}
{d\hat{\sigma}_a \over
dx}(x,Q,\mu){D}_{a/\Q}\left(\frac{z}{x},\frac{\mu}{m} \right).
\label{fac}
\end{equation}
Here the sum runs over all partons (quarks, antiquarks and gluons)
that can be produced in the  hard process and $\mu$ is the
factorization scale. The coefficient function $d\hat\sigma_a/dx$
is  the ${\overline {\rm MS}}$   renormalized
differential cross-section for producing a massless parton  $a$
\footnote{In the evaluation of the coefficient function
${d\hat{\sigma}_a/ dx}$ the heavy quark $\Q$ is considered as
massless; therefore, the sum over indexes in Eq.(\ref{fac})
includes the flavor $\Q$.}
. It is defined indirectly through the equation
\begin{equation}
{d\sigma_a \over dz}(z,Q,\epsilon) = \sum_b {d\hat\sigma_b \over
dz}(z,Q,\mu)\otimes\Gamma_{ba}(z,\mu,\epsilon), \label{facN0}
\end{equation}
where  $d\sigma_a/dz$ is the bare  energy distribution for the
parton of type $a$; the collinear divergences in this distribution
are regularized by working in $d=4-2\epsilon$ dimensions.
$\Gamma_{ab}$ are  universal collinear subtraction terms, defined
in the ${\overline {\rm MS}}$ scheme:
\begin{widetext}
\begin{eqnarray}
&& \Gamma_{ba}= \delta_{ab}\delta(1-z) -
\left({\as\over 2\pi}\right)\frac{P^{(0)}_{ab}(z)}{\epsilon}
+\left({\as\over 2\pi}\right)^2
\left[ {1\over 2 \epsilon^2}\left(
P^{(0)}_{ac}\otimes P^{(0)}_{cb}(z) +\beta_0 P^{(0)}_{ab}(z)
\right) -{1\over 2\epsilon} P^{(1)}_{ab}(z)\right],
\label{Gamma}
\end{eqnarray}
\end{widetext}
where $\as = \as(\mu)$ is the $\MSbar$
strong coupling constant, renormalized at the scale $\mu$.
The relation between the bare and the renormalized couplings
reads:
\begin{equation}
{\as^0\over 2\pi} \Se = {\as\over 2\pi}\left(1 -
{\as\over 2\pi}{\beta_0\over \epsilon} +
{\mathcal{O}}(\as^2)\right) , \label{asren}
\end{equation}
where $\Se = (4\pi)^{\epsilon}~e^{-\epsilon\gamma}$ and $\gamma$
is the Euler constant. Also, $\beta_0 = (11C_A-4T_Rn_f)/6$ is the
${\cal O}(\alpha_s^2)$ coefficient of the QCD $\beta$-function,
$C_A=3,~T_R=1/2$ are the QCD color factors, $n_f$ denotes the
number of fermion flavors (including $\Q$) and $P_{ab}^{(0,1)}$
are the time-like splitting functions \cite{furmanski}. Our
notations for the splitting functions follow
Ref.\cite{purplebook}.

The functions ${D}_{a/\Q}(x,\mu/m)$  in Eq.(\ref{fac}) are the
perturbative fragmentation functions \cite{MN}. They satisfy the
DGLAP evolution equation and can be fully reconstructed from it,
if the initial condition at a scale $\mu = \mu_0$ is known. We
denote
\begin{equation}
{ D}_{a/\Q}\left ( z,\frac{\mu_0}{m} \right )
={D}^{\rm ini}_{a} \left ( z,\frac{\mu_0}{m} \right ).
\label{D}
\end{equation}
If $\mu_0 \sim m$ is
chosen,  the initial condition ${D}^{\rm
ini}_{a}$ does not contain  large logarithms and can be
derived from fixed order perturbative  calculations.
Setting  $\mu=\mu_0$ in Eq.(\ref{fac}), we obtain:
\begin{equation}
{d\sigma^{\rm fo}_\Q \over dz}(z,Q,m) = \sum_a {d\hat{\sigma}_a
\over dz}(z,Q,\mu_0)\otimes
{ D}^{\rm ini}_{a}\left (z,\frac{\mu_0}{m} \right ),
\label{facfo}
\end{equation}
where $d\sigma^{\rm fo}$ is the fixed order cross-section  for
producing a quark $\Q$ with mass $m$, with all power corrections
${\cal O}((m^2/Q^2)^n)$, $n \ge 1$ systematically neglected. It
follows from this equation that the initial condition for the
fragmentation function can be extracted, if the heavy quark energy
spectrum is computed in a
particular process
through certain order  in $\as$. This, however, is technically inconvenient.

A more convenient, process-independent approach was suggested
recently in Refs.\cite{KL,CC}.
Since hard scattering cross-sections are insensitive to
long-distance dynamics, we may  write
\begin{eqnarray}
&&{d\sigma^{\rm fo}_\Q\over dz}(z,Q,m)=\sum_a {d
\widetilde{\sigma}_a\over dz}(z,Q,\mu_0)\otimes
\widetilde{D}_{a/\Q}\left(z,\frac{\mu_0}{m}\right) ,\label{HL1}\\
&&{d\sigma_b\over dz}(z,Q,\epsilon)=\sum_a
{d\widetilde{\sigma}_a\over dz}(z,Q,\mu_0)\otimes
\widetilde{D}^L_{a/b}(z,\mu_0,\epsilon) . \label{HL2}
\end{eqnarray}
The functions $\widetilde{D}$ and $\widetilde{D}^L$ describe
collinear radiation from massive (massless) quarks; we will show
below that they can be considered as bare fragmentation functions.
Combining Eqs.(\ref{HL1},\ref{HL2}) with
Eqs.(\ref{facfo},\ref{facN0}), we obtain:
\begin{eqnarray}
&& \widetilde{D}_{a/\Q} \left (z,\frac{\mu_0}{m} \right ) =
\sum_{b,c}\widetilde{D}^L_{a/b}(z,\mu_0,\epsilon)\otimes \left[
\Gamma(z,\mu_0,\epsilon)\right]_{bc}^{-1}\otimes \D_{c}
\left ( z,\frac{\mu_0}{m} \right ).
\label{result}
\end{eqnarray}
It follows from this equation that $\D$ can be derived once
$\widetilde{D}$ and $\widetilde{D}^L$ are known. We describe how
to compute $\widetilde{D}$ and $\widetilde{D}^L$ in the next
Section.

To conclude this Section, we  comment on the analytic structure of
$\D$. The initial condition for the perturbative fragmentation
function can be expanded in series of $\alpha_s$:
\begin{equation}
{D}^{\rm ini}_{a}\left (z,\frac{\mu_0}{m} \right )
= \sum_{n=0} \left({\as(\mu_0)\over
2\pi}\right)^n d^{(n)}_{a} \left (z,\frac{\mu_0}{m} \right ).
\label{expansionD}
\end{equation}
The results through  ${\cal O}(\alpha_s)$ are known \cite{MN}:
\begin{eqnarray}
&&d^{(0)}_a(z) = \delta_{a\Q} \delta(1-z),\nonumber\\
&&d_{a=\Q}^{(1)}\left (z,\frac{\mu_0}{m} \right )
= C_F \left[{{1+z^2}\over{1-z}}
\left( \ln\left({\mu_{0}^2}\over{m^2(1-z)^2}\right)
-1\right)\right]_+,\nonumber\\
&&d_{a=g}^{(1)}\left (z,\frac{\mu_0}{m} \right )
= T_R \left( z^2 + (1-z)^2 \right)
\ln\left( {\mu_0^2\over m^2}\right),\nonumber\\
&&d_{a\neq \Q,g}^{(1)}\left (z,\frac{\mu_0}{m} \right )= 0,
\label{DiniNLO}
\end{eqnarray}
where $C_F = 4/3$ is the QCD color factor. In this paper, we are
mostly concerned with the computation of the coefficients
$d_{\Q,~\overline\Q,~q,~\bar{q}}^{(2)}$, where $q$ stands for any
massless quark. The logarithmically enhanced  terms  in $d^{(2)}$
follow  from the DGLAP equation:
\begin{eqnarray}
d^{(2)}_a \left (z,\frac{\mu_0}{m} \right ) &=&  \left[ {
P^{(0)}_{ba}\otimes P^{(0)}_{\Q b}(z) \over 2} + {\beta_0 \over 2
} P^{(0)}_{\Q a}(z) \right] \ln^2 \left ({\mu_0^2 \over m^2}
\right )
\nonumber\\
&+& \left[ P^{(1)}_{\Q a}(z) + P^{(0)}_{ba}\otimes d_b^{(1)}(z,1)
+\beta_0 d_a^{(1)}(z,1) \right] \ln \left ({\mu_0^2 \over m^2}
\right ) + d_a^{(2)}(z,1).
\label{Dgeneral}
\end{eqnarray}
The mass-independent constant of integration $d_a^{(2)}(z,1) $ can
not be determined from the DGLAP equation and has to be computed
explicitly. This is the major goal of this paper. In the next
Section we describe how this can  be done.

\section{The collinear limit}\label{SECcollinearfact}

The idea that allows explicit computation of the functions
$\widetilde{D}^L$ and $\widetilde{D}$ and, therefore, of the
initial condition for the perturbative fragmentation function is
as follows. Both $\widetilde{D}$ and $\widetilde{D}^L$ are
introduced to describe  collinear radiation with $q_{\bot} \sim m
\leq \mu_0$. Since we are not interested in power-suppressed
contributions to the cross-section, we  can restrict ourselves to
such terms in scattering amplitudes that, upon integration over
the phase-space of final state particles, produce  ${\cal
O}(\ln^n(m)),~n \ge 1$ terms\footnote{To be precise, we are
interested in all contributions to the scattering amplitude that,
upon integration over the phase space, scale as $m^{-n\epsilon}$,
where $n$ is some integer. Upon expanding in $\epsilon$, those
terms produce {\it both} $\ln(m)$-enhanced  and $m$-independent
contributions to the energy spectrum.}. Such terms can be
identified by applying power counting arguments \cite{Ellis} in
the collinear limit for scattering amplitudes. The collinear limit
is defined as the kinematic limit where the relative transverse
momentum of two or more particles vanishes. When masses are
introduced, the collinear limit is $q_{\bot}\to 0$, $m \to 0$ and
$q_{\bot}/m = {\rm const}$ \cite{CDT}.

Throughout the paper we work in the light-cone gauge $n_\mu
A^{\mu} = 0$, $n^2 = 0$. This is necessary for the
process-independent derivation of $\D$ because in such gauges, the
${\cal O}(\ln^n(m)),~n \ge 1$ terms are produced only by diagrams
where collinear radiation is both emitted and absorbed by the same
parton. The quantum interference effects in such gauges can be
integrated over the phase-space of the final state particles in
the massless approximation. Because of that, $\D$ can be computed
from  self-energy type diagrams, integrated over the virtuality of
the incoming parton. Different cuts of such self-energy diagrams
correspond to different contributions to the fragmentation
function:  at ${\cal O}(\alpha_s^2)$, we have to deal with
two-loop virtual corrections, one-loop virtual corrections to
one-to-two splittings and, finally, one-to-three splittings.

We begin by considering the one-to-three collinear splitting since
the kinematics of the final state in this case is the most
general. We denote the four-momenta of the produced (massive or
massless) particles by $q_{1,2,3}$ and their sum by
$\widehat{p}~$:
\begin{equation}
\widehat{p}=q_1+q_2+q_3. \label{phatsum}
\end{equation}
We parameterize the collinear direction by $p$ and
use the gauge fixing vector $n$ as the complimentary light-cone
vector. We then write:
\begin{eqnarray}
q_i &=& z_ip + \beta_i {n\over (p n)} + q_{i,\bot} .
\label{sudakov}
\end{eqnarray}
The components $\beta_i$ are found from
the on-shell conditions $q_i^2=m_i^2$:
\begin{equation}
\beta_i = {-q^2_{i,\bot}+m_i^2-z_i^2m_p^2 \over 2z_i}~~,~~~ i=1,2,3
~, \nonumber
\end{equation}
with $p^2=m_p^2$. We are interested in the case when
$\widehat{p},p$ and $n$ belong to the same hyperplane in the
$d$-dimensional space\footnote{More general configuration has been
considered in \cite{CG}.}. Then:
\begin{equation}
z_1+z_2+z_3 = 1~ ,~~ q_{1,\bot}+q_{2,\bot}+ q_{3,\bot} = 0 .
\label{constr}
\end{equation}
Using the above equations, we express the momentum
$\widehat{p}$ through $p$ and $n$:
\begin{equation}
\widehat{p} = p + {1\over 2}(\widehat{p}^2-m_p^2){n\over (pn)},
\label{phat}
\end{equation}
where:
\begin{eqnarray}
\widehat{p}^2-m_p^2 = \frac{2(pq_2)+ 2(pq_3)-2(q_2q_3) +
m_1^2-m_2^2-m_3^2-m_p^2 }{z}. \nonumber
\end{eqnarray}
Our notations are such that $q_1$ always denotes the momentum of
the heavy quark in the final state, whose energy is measured. For
that reason we often write $z$ instead of $z_1$, to denote its
energy fraction. Eqs.(\ref{sudakov},\ref{constr})  imply the
relation:
\begin{equation}
z = 1 - {(nq_2)\over (pn)} -{(nq_3)\over (pn) }, \label{z}
\end{equation}
which is nothing but the energy conservation condition.

Although we have only discussed the kinematics of the one-to-three
splitting, the kinematics of the one-to-two splitting can be
obtained as a particular case. For this, it is sufficient to  set
the momentum $q_3$ and the mass $m_3$ to zero in the equations
above.

In QCD, factorization properties of the cross-sections, as in
Eqs.(\ref{HL1},\ref{HL2}), can be traced back to the factorization
properties of both the phase-space and the matrix elements in the
collinear kinematics. We illustrate  those properties by
considering the tree level amplitudes and then generalize these
considerations to include virtual corrections.

Consider a hard scattering process characterized by some scale $Q
\gg m$. Suppose that $n+2$ partons with momenta
$k_1,\dots,k_{n-1},q_1,q_2,q_3$ are produced.  We consider momenta
$k_1,\dots,k_{n-1}$ as  non-exceptional, while momenta
$q_1,q_2,q_3$ are collinear, as described above. First, we discuss
the phase-space factorization.

We denote the phase-space of $n+2$ partons as
${\rm dPS}^{(n+2)}(k_1,\dots,k_{n-1},q_1,q_2,q_3)$.
In the limit when $q_1,q_2$ and $q_3$ become collinear
$q_1+q_2+q_3=p+{\mathcal{O}}(q_\bot)$, the $(n+2)$-particle phase
space factorizes:
\begin{eqnarray}
{\rm dPS}^{(n+2)}(k_1,\dots,k_{n-1},q_1,q_2,q_3) =
{\rm dPS}^{(n)}(k_1,\dots,k_{n-1},p)~
{\rm d}\Phi^{\rm coll}(q_2,q_3).
\label{PSfact}
\end{eqnarray}
We use the following notations:
\begin{equation}
{\rm d}\Phi^{\rm coll}(q_2,q_3) = {1\over z} [dq_2;m_2][dq_3;m_3] ,
\label{PSfact2}
\end{equation}
and $[dq;m_q]$ is the $d$-dimensional one-particle phase space:
\begin{equation}
[dq;m_q] = {d^dq\over (2\pi)^{d-1}}\delta^+(q^2-m_q^2) . \label{onePS}
\end{equation}
To derive Eq.(\ref{PSfact}), we neglect the ${\cal O}(q_\bot, m)$
difference between $\hat p$ and $p$ in ${\rm dPS}^{(n)}$ and
integrate over $q_1$ using Eq.(\ref{phatsum}).

We now consider  factorization properties  of the tree-level matrix
element. In the massless case, they were described  in
\cite{CG,CampGlover}. Below we generalize those results to the
case of non-zero mass $m$. We consider massive quarks produced in
the fragmentation of  a quark-like  parton of either the
same or different flavor. The case of a gluon fragmentation into a
heavy quark can be dealt with in a similar way.

We already mentioned that, in physical gauges, the relevant
contributions in the collinear limit come from diagrams where the
same parton emits and absorbs collinear radiation. Taken in
conjunction with power counting arguments, this observation can be
used to simplify tree-level matrix elements in the collinear limit
and derive the factorization formula. Consider a process with
$n+2$ particles in the final state described by the  matrix
element $M^{(n+2)}(k_1,\dots,k_{n-1},q_1,q_2,q_3)$. In physical
gauges, the collinear splitting $\widehat{p}\to q_1+q_2+q_3$
decouples from the rest of the process, so that the squared,
spin-averaged amplitude $\vert M^{(n+2)}\vert^2$ can be written as
\cite{CG}:
\begin{eqnarray}
&& \vert M^{(n+2)}(k_1,\dots,k_{n-1},q_1,q_2,q_3) \vert^2 =
 {\cal{M}}^{(n)}(k_1,\dots,k_{n-1},\widehat{p})_{\beta, \alpha}
V^{\rm coll}_{\alpha,\beta}(\widehat{p},q_2,q_3). \label{Mfact}
\end{eqnarray}
Here $\alpha,\beta$ are spinor indices. If, in the collinear limit,
$V^{\rm coll}$ can be written as
\begin{equation}
V_{\alpha,\beta}^{\rm coll}= \not{\!p}_{\alpha,\beta}~ W
+{\mathcal{O}}(q_\bot), \label{VtoW}
\end{equation}
we can rewrite Eq.(\ref{Mfact}) to make the factorization of the
matrix element explicit:
\begin{eqnarray}
&& \vert M^{(n+2)}(k_1,\dots,k_{n-1},q_1,q_2,q_3)\vert^2 =
\vert M^{(n)}(k_1,\dots,k_{n-1},p)\vert^2~ W(n,\widehat{p},q_2,q_3)
+{\mathcal{O}}(q_\bot).
\label{MfactW}
\end{eqnarray}
Here $|M^{(n)}(....,p)|^2$ is the amplitude squared for producing
$n$ on-shell particles with non-exceptional momenta; the scalar
function $W$ contains all the information about the collinear
splitting.

We now explain why $V^{\rm coll}$ can be written as in
Eq.(\ref{VtoW}). Because the collinear splitting $\widehat{p}\to
q_1+q_2+q_3$ decouples from the rest of the process, $V^{\rm
coll}$ is obtained as the sum squared of all possible diagrams that
describe a transition of an off-shell parton with momentum
$\widehat{p}$ into three particles with momenta $q_{1,2,3}$.
The propagator of  the off-shell parton $\widehat{p}$
is included in $V^{\rm coll}$. In the massless case the
matrix $V^{\rm coll}(n,\widehat{p},q_2,q_3)$
can be written as \cite{CG}:
\begin{equation}
V^{\rm coll} = \left({\mu_0^{2\e}\over
{\widehat{p}}^2}\right)^2\left(\sum_{i=1}^3 A_i\not{\!q}_i + B
{\not{\!n}\over (pn)}\widehat{p}^2\right) . \label{V}
\end{equation}
The functions $A_i$ and $B$ are dimensionless  functions that
depend on the  scalar products $(q_i q_j)$. To determine the
behavior of the amplitude $V^{\rm coll}$ in the collinear limit,
we rescale the transverse momenta:
\begin{equation}
q_{i,\bot} \to \kappa q_{i,\bot}\label{rescale}
\end{equation}
and consider the limit $\kappa\to 0$. Since
$(q_i q_j) \sim {\mathcal{O}}(q_\bot^2)$,
the functions $A_{1-3}$ and $B$ remain
invariant under that rescaling. As follows from
Eq.(\ref{sudakov}), the matrices $ \not{\!q}_i $ transform as:
$\not{\!q}_i \to z_i\not{\!p} +  {\mathcal{O}}(\kappa)$.
Therefore, after the rescaling Eq.(\ref{rescale}), the amplitude
$V^{\rm coll}$ becomes:
\begin{equation}
V^{\rm coll} = \kappa^{-4}\left({\mu_0^{2\e}\over
\widehat{p}^2}\right)^2\left(\sum_{i=1}^3 A_iz_i\right)
\not{\!p}  + {\mathcal{O}}(\kappa^{-3}). \label{Vkappa}
\end{equation}
Because  the collinear phase space  Eq.(\ref{PSfact2}) transforms
as:
\begin{equation}
{\rm d}\Phi^{\rm coll}(q_2,q_3) \to \kappa^{4-4\epsilon}
{\rm d}\Phi^{\rm coll}(q_2,q_3) ,
\label{collmeasurescale}
\end{equation}
under the rescaling Eq.(\ref{rescale}), it follows  that in the
collinear limit $\kappa \to 0$ only the first term in
Eq.(\ref{Vkappa}) gives non-vanishing  contribution after the
phase-space integration. This proves that $V^{\rm coll}$ has the
form shown in Eq.(\ref{VtoW}).  We can extract the function $W$
from  $V^{\rm coll}$ by applying the projection:
\begin{equation}
W = {{\rm Tr}\left[\not{\!n} V^{\rm coll}\right] \over 4(pn) } .
\label{projector}
\end{equation}

We now generalize this result to the case when at least one of the
final state particles is a quark with mass $m$. The mass $m$ is
considered to be of the order of the transverse momenta of the
collinear particles. We  follow the same line of reasoning that
leads to Eq.(\ref{V}) allowing, however, for non-vanishing masses
in the initial and final states of the collinear splitting
process. Eq.(\ref{V}) generalizes to:
\begin{equation}
V^{\rm coll} = \left(\frac{\mu_0^{2\e}}{
{\widehat{p}}^2-m_{in}^2}\right)^2 \left(\sum \limits_{i=1}^{3}
A^{(m)}_i\not{\!q}_i + B^{(m)} {\not{\!n}\over (pn)} + C m\right)
. \label{Vm}
\end{equation}
In Eq.(\ref{Vm}), $m_{in}$ is the mass of the parton that
initiates the collinear splitting. The functions
$A^{(m)}_i,~B^{(m)}$ and $C$ are scalar functions of dimension
zero, two and zero respectively; they depend on $(q_iq_j)$ and
$m^2$. These functions possess a regular $m \to 0$ limit.  In the
collinear limit, when  both the transverse momenta and the mass
$m$ are rescaled simultaneously,
\begin{eqnarray}
q_{i,\bot} \to \kappa~ q_{i,\bot},~~~~
m \to \kappa~ m,   \label{rescalem}
\end{eqnarray}
the scaling properties of the functions $A^{(m)}_i,~B^{(m)}$ and
$C$ coincide with their mass dimension. Therefore, under the
rescaling Eq.(\ref{rescalem}), the amplitude $V^{\rm coll}$
behaves as:
\begin{equation}
V^{\rm coll} = \kappa^{-4}\left({\mu_0^{2\e}\over
\widehat{p}^2-m_{in}^2}\right)^2 \left(\sum_{i=1}^3 A^{(m)}_iz_i
\right)\not{\!p} + {\mathcal{O}}(\kappa^{-3}). \label{Vkappam}
\end{equation}
This result implies  that also for the $m\neq 0$ case,  the function
$W$ is given by Eq.(\ref{projector}).
Finally, assembling all the pieces, we find
that the function $\widetilde{D}(z)$ in Eq.(\ref{HL1}),
can be written as:
\begin{eqnarray}
\widetilde{D}(z) = {1\over z} \int [dq_2;m_2][dq_3;m_3] {{\rm
Tr}\left[\not{\!n} V^{\rm coll}(\widehat{p},q_2,q_3,n;m)\right]
\over 4(pn) } \delta\left(1 -z - {(nq_2)\over (pn)} -{(nq_3)\over
(pn) } \right) . \label{Dresult}
\end{eqnarray}
Although we have derived Eq.(\ref{Dresult}) for the one-to-three
splitting contribution to  $\D_{a}(z,\mu_0/m)$, similar expression
is valid
when virtual corrections are included. In addition,
Eq.(\ref{Dresult}) can be used to derive the function
$\widetilde{D}^L$ after the mass $m$ is set to zero everywhere.

We now discuss how to perform phase space integrations in
Eq.(\ref{Dresult}). The approach proposed in \cite{CC,KL} requires
integration over the transverse momentum up to the scale
$\mu_{\perp} = \mu_0$ for both $\widetilde{D}$ and
$\widetilde{D}^L$. While this allows a simple calculation of the
fragmentation function at ${\cal O}(\alpha_s)$,  this approach
becomes impractical  in higher orders of perturbation theory. It
is possible to simplify the calculation considerably by realizing
that the limit $\mu_{\perp} \to \infty$ can be taken when the {\it
difference} of  massive and massless functions $\widetilde D$ is
considered. The technical simplification associated with this is
twofold.  First, all higher order QCD effects in the massless
function $\widetilde{D}^L$ vanish, because all the integrals that
contribute there become scaleless. We obtain:
\begin{equation}
\widetilde{D}^L_{a/b}(z,\mu_0,\epsilon) = \delta_{ab}\delta(1-z).
\label{DLresult}
\end{equation}
Second, in the massive case, only single-scale  integrals have to
be computed. Using Eq.(\ref{DLresult}), we obtain a simple
expression for the initial condition of the perturbative
fragmentation function:
\begin{equation}
\D_{a}\left (z,{\mu_0 \over m} \right) = \sum_b
\Gamma_{ab}(z,\mu_0) \otimes \widetilde{D}_{b/\Q}\left (z,{\mu_0
\over m} \right). \label{DHresult}
\end{equation}
From this it follows, that $\widetilde{D}_{b/\Q}$ is
 the  bare fragmentation function for the massive quark,
whereas $\D$ represents its collinear renormalized version.

As we mentioned earlier, there are three contributions to
$\widetilde{D}_{b/\Q}(z,\mu_0,m)$, that have to be considered:
double-virtual, real-virtual and double-real. In all these cases
the amplitude $V^{\rm coll}$ can be constructed by applying
conventional Feynman rules to describe the transition of a parton
with momentum $\widehat{p}$ to the final state of up to three
particles with momenta $q_{1,2,3}$.  Virtual corrections should
also be included, when appropriate. For each of these three cases,
the function $W$ is obtained using the same projector as in
Eq.(\ref{DHresult}). The only unusual feature of the amplitudes
that contribute to $V^{\rm coll}$
is that the spinor, that describes the initial state  parton, has to be
replaced  with the propagator of the same parton with the
off-shell momentum $\widehat{p}$.  $V^{\rm coll}$ is then given by
the square of the corresponding matrix element. Standard symmetry
factors apply in this case as well. In particular, to evaluate
$\widetilde{D}$ one has to sum over spins and colors of the final
state and to average over colors in the initial state.
The ultraviolet renormalization is performed in a standard
fashion. One has to renormalize the QCD coupling constant, the
quark mass and external quark and gluon fields. Since we work in the
light cone gauge, we find it convenient to compute the quark and
gluon field renormalization constants by computing  diagrams with
self-energy  insertions  to external quark and gluon lines.

Finally, we comment on  the double virtual contributions to the
fragmentation function. Such contributions describe the one-to-one
splitting and are  proportional to $\delta(1-z)$. Unfortunately,
computing these contributions requires dealing with well-known
complications of light-cone gauges. For this reason, we decided to fix
the ($\e$-dependent) coefficient of $\delta(1-z)$ using
the  fermion number conservation sum rule:
\begin{equation}
\int_0^1 dz \left( \D_{\Q/\Q}(z) - \D_{\overline\Q/Q} \right) = 1.
\label{FNC}
\end{equation}
We do not encounter, however, any spurious light-cone
singularities in the diagrams with one virtual loop. All
singularities originating from those diagrams are consistently
regulated with dimensional regularization.

To perform  integrations in   phase-space and loop integrals we
proceed as in  \cite{AM,ADM,ADMP}, where further details of the
method can be found. The idea is to map all  non-trivial
phase-space integrals to loop integrals and use standard multiloop
methods \cite{tkachov}, such as integration-by-parts and
recurrence relations, to reduce all the   phase-space and loop
integrals that have to be  evaluated, to a few master integrals.
For the reduction to master integrals we use the algorithm
\cite{Laporta} implemented in \cite{babis}. All algebraic
manipulations have been performed using Maple \cite{MAPLE} and
Form \cite{FORM}. We now discuss the calculation of the initial
condition $\D$ through ${\cal O}(\alpha_s)$ which allows us to
illustrate the details of the method by considering a simple
example.

\section{$\D$ at NLO}\label{SecNLO}

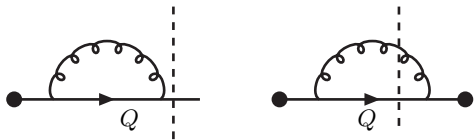
\begin{figure}
\begin{picture}(200,40)(0,-20)
%\SetColor{Black}
\SetWidth{1}
\Vertex(20,10){3}
\ArrowLine(20,10)(90,10)
\GlueArc(55, 10)(20,0, 180){2}{6}
\DashLine(80,45)(80,-5){3}
\Vertex(120,10){3}
\ArrowLine(120,10)(190,10)
\GlueArc(155, 10)(20,0, 180){2}{6}
\DashLine(165,45)(165,0){3}
\Vertex(190,10){3}
\put(60,0){$Q$}
\put(150,0){$Q$}
\end{picture}
\caption{Diagrams that contribute to perturbative fragmentation
of a heavy quark  $Q \to Q + X$
at ${\cal O}(\alpha_s)$. The dashed vertical line
indicates the intermediate state that has to be considered.}
\end{figure}
In this Section we compute the initial condition for the
fragmentation function at NLO through ${\cal O}(\epsilon)$. Such
terms are needed for the computation of $\D$ at order ${\cal
O}(\alpha_s^2)$. The function $V^{\rm coll}$ for $\Q \to \Q + g$
splitting is obtained by considering the diagrams shown in Fig.1.
To facilitate the comparison with the literature, we express the
function $W$, Eq.(\ref{projector}), through the so-called
splitting function $P$ \cite{CG}:
\begin{equation}
W = \left( { 8\pi\as\mu_0^{2\e} \over \widehat{p}^2-m^2 }\right)
P. \label{P}
\end{equation}
Introducing auxiliary vector $\widehat{n}$:
\begin{equation}
{n \over (pn)} = (1-z)\widehat{n} , \label{nhat}
\end{equation}
and using the projector Eq.(\ref{projector}), we obtain the
splitting function:
\begin{equation}
P^{(1)} = C_F\left( 2{(pn)\over (qn)} + (1-\e){(qn)\over (pn)} -2
-m^2 {(pn)-(qn)\over (pq)(pn)}  \right) .
\label{P1}
\end{equation}

Combining Eqs.(\ref{z},\ref{nhat}), we derive the real emission
contribution  to the fragmentation function $\widetilde{D}$:
\begin{eqnarray}
\widetilde{D}^{(r)}(z) &=&
\left ( {\as^0\over 2\pi} \right ){8\pi^2\mu_0^{2\e}\over
1-z} \int [dq;0] {P^{(1)} \over (pq) }
\delta\left (1 - (\widehat{n}q)\right ) .
\label{D1r}
\end{eqnarray}
Because of the constraint $\delta\left (1 - (\widehat{n}q)\right )$
in Eq.(\ref{D1r}), the splitting function simplifies:
\begin{equation}
P^{(1)} = C_F\left({2\over 1-z} -1-z -\e(1-z) - z{m^2\over (pq)}
\right). \label{P1z}
\end{equation}
The result for $P^{(1)}$ in Eq.(\ref{P1z}) agrees with
Ref.\cite{CDT}. When $P^{(1)}$ is used in Eq.(\ref{D1r}) we
observe that two integrals have to be evaluated. However, those
integrals are not independent; an algebraic relation between them
can be found using the method of \cite{AM,ADM,ADMP}. As a result,
$\widetilde{D}^{(r)}(z)$ can be expressed through a single
``master'' integral:
\begin{eqnarray}
&& I^{(1)} = \int { d^dq \over (pq) }
\delta\left[q^2\right]\delta\left[1 - (\widehat{n}q)\right]
= (\pi)^{1-\e} \Gamma(\e)m^{-2\e}(1-z)^{1-2\e}.
\label{abc}
\end{eqnarray}

We finally obtain the real emission contribution to the
fragmentation function:
\begin{equation}
\widetilde{D}^{(r)}\left (z,\frac{\mu_0}{m} \right ) =
\left ( {\as^0\over
2\pi} \right )
C_F \left({4\pi \mu_0^2\over m^2}\right)^\e
\frac{(1-\e)\Gamma(\e)(1+z^2)}{(1-z)^{1+2\e}}, \label{D1rfin}
\end{equation}
valid to all orders in $\epsilon$. To expand this result  in $\e$,
we use
\begin{equation}
(1-z)^{-1+a\e} = \frac{\delta(1-z)}{a\e} +\sum_{n\geq 0}
{(a\e)^n\over n!} \left [ {\ln^n(1-z) \over 1-z} \right]_+ .
\label{plusid}
\end{equation}

Virtual corrections are derived  by considering self-energy
diagrams in the light-cone gauge. The result reads:
\begin{widetext}
\begin{eqnarray}
&& \widetilde{D}^{(v)}\left (z,\frac{\mu_0}{m} \right ) =
\left ( {\as^0\over 2\pi} \right )S_\e
C_F\delta(1-z)\Bigg\{ {1\over \e^2} +{1\over
\e}\left[\ln\left({\mu_0^2\over m^2}\right)+{1\over 2}\right] +
{1\over 2}\ln^2\left({\mu_0^2\over m^2}\right)
+ {1\over 2}\ln\left({\mu_0^2\over m^2}\right)+{\pi^2\over 12}+2
\nonumber \\
&& +\e\left[ {1\over 6}\ln^3\left({\mu_0^2\over
m^2}\right)+{1\over 4}\ln^2\left({\mu_0^2\over m^2}\right)
+\left({\pi^2\over 12}+2\right)\ln\left({\mu_0^2\over
m^2}\right) +{\pi^2\over 24} - {\zeta(3)\over 3} + 4 \right]
\Bigg\}+{\mathcal{O}}(\e^2) . \label{MNvirt}
\end{eqnarray}
\end{widetext}

Combining  Eqs.(\ref{D1rfin}-\ref{MNvirt}), we arrive at the bare
perturbative fragmentation function through ${\cal O}(\alpha_s)$:
\begin{widetext}
\begin{eqnarray}
&& \widetilde{D}_{\Q/\Q} \left ( z,\frac{\mu_0}{m} \right ) =
\delta(1-z) + \left ( {\as^0\over 2\pi} \right )S_\e
C_F\Bigg\{ {1\over\e}\left [{1+z^2\over 1-z}\right]_+ + \left[
{1+z^2\over 1-z}\left( \ln\left({\mu_0^2\over
m^2}\right)-2\ln(1-z)-1\right)\right]_+ \nonumber\\
&& + \e\left[ {1+z^2\over 1-z}\left( {1\over
2}\ln^2\left({\mu_0^2\over m^2}\right) - \ln\left({\mu_0^2\over
m^2}\right)\left(2\ln(1-z)+1\right)
+ 2\ln^2(1-z)+2\ln(1-z) +{\pi^2\over 12} \right)
\right]_+ + {\mathcal{O}}(\e^2) \Bigg\}.
\label{D1}
\end{eqnarray}
\end{widetext}
As expected from the fermion number conservation condition,
Eq.(\ref{FNC}), the  integral of
$\widetilde{D}_{\Q/\Q}(z,\mu_0/m)$ over $z$ equals $1$ and does
not receive any corrections at order ${\cal O}(\alpha_s)$.
Applying collinear renormalization Eq.(\ref{DHresult}) and
substituting $\as^0 S_\e \to \as$ in Eq.(\ref{D1}), we observe
that the $1/\epsilon$ term in Eq.(\ref{D1}) cancels and the
remaining  terms are not modified. Therefore $\D$ can be read off
from Eq.(\ref{D1}) by simply omitting the $1/\epsilon$ term. The
${\cal O}(\e^0)$ part of $\D$ coincides with the result of
Ref.\cite{MN}, while the ${\cal O}(\e)$ part is new.

The gluon fragmentation $g \to Q + \overline{Q}$ can be treated in
a similar way. The real emission contribution reads:
\begin{equation}
\widetilde{D}_{g/\Q}\left (z,\frac{\mu_0}{m} \right ) =
{\as^0\over 2\pi} {4(2\pi)^{-1+2\e}\mu_0^{2\e}\over z(1-z)} \int
[dq;m] {P_{g/\Q}\over \widehat{p}^2} \delta(1-(\widehat{n}q)) .
\label{D1g}
\end{equation}
Evaluation of the splitting function $P_{g/\Q}$ is described
in \cite{CC}.  We derive:
\begin{equation}
P_{g/\Q} = T_R\left( 1-2z{1-z\over 1-\e} + {z\over 1-\e}{m^2\over
(pq)} \right), \label{P1g}
\end{equation}
which agrees with the result in Ref.\cite{CDT}. There are two
integrals of the type:
\begin{equation}
J^{(a)} = \int { d^dq \over (pq)^a }
\delta\left[q^2-m^2\right]\delta\left[1 - (\widehat{n}q)\right]
~~,~~a=1,2~, \label{abcg}
\end{equation}
that have to be computed, but only one of them is independent. We
obtain:
\begin{equation}
J^{(1)} = (\pi)^{1-\e} \Gamma(\e)m^{-2\e}(1-z) . \nonumber
\end{equation}
This allows us to derive the gluon-initiated contribution to the
initial condition for the perturbative fragmentation function at
${\cal O}(\as)$:
\begin{eqnarray}
&& \widetilde{D}_{g/\Q}\left (
z,\frac{\mu_0}{m} \right ) = \left ( {\as^0\over 2\pi} \right )S_\e T_R
\left(z^2+(1-z)^2 \right)\Bigg\{ {1\over\e} +
\ln\left({\mu_0^2\over m^2}\right)
+ \e\left[ {1\over 2}\ln^2\left({\mu_0^2\over m^2}\right)
+{\pi^2\over 12} \right]\Bigg\} + {\mathcal{O}}(\e^2).
\label{D1gr}
\end{eqnarray}
The $1/\e$ pole cancels after collinear renormalization and
 the known result  for $\widetilde{D}_{g/\Q}$ \cite{MN}
is
recovered in the
limit $\e\to 0$. Having explained out method by considering a simple
example, we now present the results for $\D$ through ${\cal
O}(\alpha_s^2)$.

\section{$\D$ at NNLO}

At NNLO, large number of processes contribute,   yet, all of
them can be treated using the methods described above. Before we
present the results for $\D$ through ${\cal O}(\alpha_s^2)$, we
briefly mention some technical details that we find peculiar.

Several sub-processes contribute at ${\cal O}(\alpha_s^2)$. It is
convenient to split them according to their contributions to
various components of $\D_a$.  The real emission contributions to
$\D_\Q$ are $\Q \to \Q gg, \Q q \bar q, \Q \Q \overline{\Q}$,
while for $\D_{\overline{\Q}}$ and $\D_{q,\bar q}$ the real
emission sub-processes are $\overline{\Q} \to \Q \overline{\Q}
\overline{\Q}$ and $q(\bar q) \to \Q q(\bar q)\overline{\Q}$,
respectively. Some of the contributing diagrams are shown on
Fig.2.

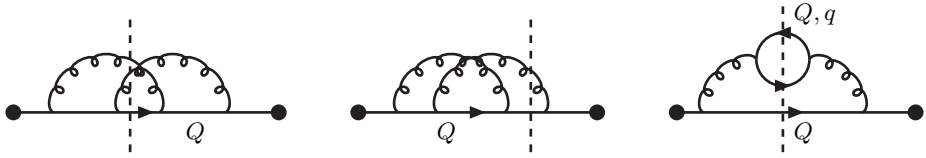
\begin{figure}
\begin{center}
\begin{picture}(200,40)(0,-20)
%\SetColor{Black}
\SetWidth{1}
\Vertex(-70,10){3}
\ArrowLine(-70,10)(30,10)
\GlueArc(-35, 10)(20,0, 180){2}{6}
\GlueArc(-10, 10)(20,0,180){2}{6}
\DashLine(-26,45)(-26,-5){3}
\Vertex(30,10){3}
\Vertex(60,10){3}
\ArrowLine(60,10)(150,10)
\GlueArc(95, 10)(20,0, 180){2}{6}
\GlueArc(110, 10)(20,0,180){2}{6}
\DashLine(125,45)(125,-5){3}
\Vertex(150,10){3}
\Vertex(180,10){3}
\ArrowLine(180,10)(270,10)
\GlueArc(230, 10)(20,0, 90){2}{3}
\GlueArc(210, 10)(20,90, 180){2}{3}
\ArrowArc(220,30)(10,0,180)
\ArrowArc(220,30)(10,180,360)
\DashLine(220,50)(220,-5){3}
\Vertex(270,10){3}
\put(-5,0){$Q$}
\put(90,0){$Q$}
\put(225,45){$Q,q$}
\put(225,0){$Q$}
\end{picture}
\end{center}
\caption{Examples of diagrams that contribute to perturbative fragmentation
of a heavy quark  $Q \to Q + X$
at ${\cal O}(\alpha_s^2)$. The dashed vertical line
indicates the intermediate state that has to be considered.}
\end{figure}

Because the number of contributing processes is large,
there are many possibilities of mass assignments for  particles
in the final state, relevant for the calculation
of $\D$. We have to consider all of those cases separately.
As we explained in Section III, we require
 the tree-level splitting amplitudes in the
collinear limit when massive quarks are present. Those results
are not available in the literature; yet, a useful cross-check
of our results is obtained once the limit $m \to 0$ is taken.
In that limit the  splitting amplitudes derived in this paper
coincide with those in Ref.\cite{CG}.

We now present the final results for the fermion initiated
contributions to the initial condition of the perturbative
fragmentation function at order $\as^2$. We give the results for
the coefficients $d^{(2)}_{a}(z,\mu_0/m)$, introduced in
Eq.(\ref{expansionD}).   Our
results contain polylogarithmic functions up to
rank three. These functions are defined through:
\begin{equation}
\Li_n(z)=\int_0^z{\Li_{n-1}(x)\over x} dx , ~~ \Li_1(z)=-\ln(1-z) ,
~~~S_{1,2}(z)={1\over 2}\int_0^z{\ln^2(1-x)\over x} dx .
\nonumber
\end{equation}

We begin with  the simplest case of a light (anti)quark-initiated
fragmentation process. Through ${\cal O}(\alpha_s^2)$, the
two contributions coincide:
\begin{equation}
d^{(2)}_{q}\left(z,{\mu_0\over m}\right) =
d^{(2)}_{\overline{q}}\left(z,{\mu_0\over m}\right). \nonumber
\end{equation}
We write:
\begin{equation}
d^{(2)}_{q}=C_F T_R F^{(C_FT_R)}_{q},
\end{equation}
where
\begin{widetext}
\begin{eqnarray}
&& F^{(C_FT_R)}_{q} =
\left \{ (1+z)\ln(z)+{(1-z)(4z^2+7z+4)\over 6z}\right \}L^2
+ \left \{ (1+z)\ln^2(z)-{8z^2 +27z+15 \over 3}\ln(z)
\right. \nonumber \\
&& \left. -{4(1-z)(14z^2+23z+5)\over 9z}\right \} L
+  \left( 2(1+z)\ln(z)+{(1-z)(4z^2+7z+4)\over 3z}\right)
\left( \Li_2(z)+\ln(1-z)\ln(z)\right)
\nonumber \\
&+&  4(1+z)\left[\Li_3(z)-\ln( z)\Li_2(z)+{1\over
24}\ln^3(z)-{1\over 2}\ln(1-z)\ln^2(z)-\zeta(3)\right] -  {15 +
27z + 8z^2 \over 12} \ln^2(z)
\nonumber\\
&+& \left({16\over 3}+{56\over 3}z +{56\over 9}z^2 \right)\ln(z)+
{7(1-z)(16+133z+88z^2)\over 54z},
\end{eqnarray}
\end{widetext}
where $L=\ln(\mu_0^2/m^2)$.

\vspace*{1cm}
Next, we present the heavy antiquark-initiated contribution
$d^{(2)}_{\overline{\Q}}$. We write:
\begin{equation}
d^{(2)}_{\overline{\Q}}
= C_F\left(C_F-{C_A\over 2}
\right) F^{(C_A,C_F)}_ {\overline{\Q}}
+ C_FT_R F^{(C_FT_R)}_ {\overline{\Q}}.
\nonumber
\end{equation}
The functions $F^{(C_A,C_F)}_ {\overline{\Q}}$ and
$F^{(C_FT_R)}_ {\overline{\Q}}$ read:
\begin{widetext}
\begin{eqnarray}
&& F^{(C_A,C_F)}_ {\overline{\Q}}  = \left \{ {1+z^2\over 1+z}
\left ( -4\Li_2(-z) +\ln^2(z)-4 \ln(1+z)\ln(z)
-\frac{\pi^2}{3}\right )
+2(1+z)\ln(z) +4 (1-z) \right \} L\nonumber\\
&+& {1+z^2\over 1+z} \left (
4 \Li_3(z)
- 2\Li_3(-z)+12S_{1,2}(-z)+{\ln(z)^3\over 2}
-2\ln(z) \Li_2(z)  - 2\ln(z)\Li_2(-z)
\right. \nonumber \\
&+& \left.  12 \ln(1+z)\Li_2(-z)
-3\ln(1+z)\ln(z)^2 + 6 \ln^2(1+z) \ln(z)
-{2 \pi^2 \over 3 }\ln(z)-7\zeta(3)
+ \pi^2\ln(1+z)
\right )\nonumber\\
&-& {4(12z^2+3z^4+8z^3+1+12z)\over 3(1+z)^3} \Li_2(z)
+{2(3z^2-1-10z)\over 3(1+z)}\Li_2(-z)
- {z(-3z+3+z^2+3z^3)\over 3(1+z)^3}\ln^2(z)
\nonumber\\
&-& {4(12z^2+3z^4+8z^3+1+12z)\over 3(1+z)^3}\ln(1-z)\ln(z) +
{2(3z^2-1-10z)\over 3(1+z)}\ln(1+z)\ln(z)
\nonumber\\
&&+{-30z^2+15z^4-9-28z+36z^3\over 6(1+z)^3}\ln(z)
+{30z^2+28z^3+15z^4+36z +3\over 18(1+z)^3}\pi^2
+{45z^3-29z+29z^2-45\over 6(1+z)^2} ,
\nonumber\\
\end{eqnarray}
\end{widetext}
and
\begin{widetext}
\begin{eqnarray}
&& F^{(C_FT_R)}_ {\overline{\Q}}=
\left \{ (1+z)\ln(z)-{ 4z^2+3z-3\over 6}+{2\over 3z}\right \}L^2
+\left \{ (1+z)\ln^2(z)-{ 8z^2+27z+15\over 3}\ln(z)
\right. \nonumber \\
&& \left.
+{56z^2+36z-72\over 9}-{20\over 9z} \right \} L
+ (1+z)\left(   4\ln(z)\Li_2(z) +8\ln(z)\Li_2(-z)
-8\Li_3(z)-16\Li_3(-z)+{1\over 6}\ln^3(z)-4\zeta(3) \right)\nonumber\\
&+& {4(z-1)(z^2+4z+1)\over 3z}
\left( \Li_2(z) +2\Li_2(-z) +\ln(1-z)\ln(z) +2\ln(1+z)\ln(z) \right)
-{8z^2+27z+15\over 12}\ln^2(z)
\nonumber\\
&&+{20z^5+351z+48+147z^4+489z^3+713z^2 \over 9(1+z)^3}\ln(z)
-{400z^4-963z+1331z^3-819+499z^2\over 54(1+z)^2}+{56\over 27 z}  .
\end{eqnarray}
\end{widetext}

\vspace*{1cm}
Finally, we present the heavy quark-initiated contribution
$d^{(2)}_{\Q}(z,\mu_0/m)$. We write:
\begin{eqnarray}
d^{(2)}_{\Q}\left(z, {\mu_0\over m}  \right)
&=& C_F^2 F^{(C_F^2)}_ {\Q}
+C_AC_F F^{(C_AC_F)}_ {\Q}
+ C_FT_R  F^{(C_FT_R)}_ {\Q} +
C_FT_Rn_l F^{(C_FT_Rn_l)}_ {\Q},
\label{d2Qresult}
\end{eqnarray}

where $n_l = n_f -1 $ is the number of massless flavors. The
functions $F_\Q$ read:

\begin{widetext}
\begin{eqnarray}
&& F^{(C_F^2)}_ {\Q}  =
\left\{
\left({9\over 8} -{\pi^2\over 3}\right)\delta(1-z)
+4\left[{\ln(1-z)\over 1-z}\right]_+ +3\left[{1\over 1-z}\right]_+
 -{1+3z^2\over 2(1-z)}\ln(z)
-2(1+z)\ln(1-z) \right. \nonumber\\
&-& \left. {5+z\over 2}\right \} L^2
+\left\{
\left({27\over 8}+{\pi^2\over 6} -2\zeta(3) \right)\delta(1-z)
-12\left[{\ln^2(1-z)\over 1-z}\right]_+
-14\left[{\ln(1-z)\over 1-z}\right]_+
+ \left (1 + { 4\pi^2\over 3} \right )
\left[{1\over 1-z}\right]_+ \right.\nonumber\\
&+& \left. (1+z) \left (2\Li_2(z) -\pi^2 \right )
-{3 + 5z^2\over 2(1-z)}\ln^2(z)
+  {6(1+z^2) \over (1-z) } \ln(1-z)\ln(z)
+{(4z^2+4z-1)\over 1-z} \ln(z)
\right. \nonumber \\
&& \left. +6(1+z)\ln^2(1-z)+(11+3z)\ln(1-z) +{9z-11\over 2}\right
\} L +\delta(1-z)\left( {241\over 32}+{7\pi^2\over
12}-2\pi^2\ln(2) -{5\pi^4\over 36}+{13 \zeta(3)\over 2}\right)
\nonumber\\
&+&
8\left[{\ln^3(1-z)\over 1-z}\right]_+
+12\left[{ \ln^2(1-z)\over 1-z}\right]_+
 -\left(4+{8\pi^2\over 3}\right)\left[{\ln(1-z)\over 1-z}\right]_+
-\left( 4+{4\pi^2\over 3}
-16\zeta(3)\right)\left[{1\over 1-z}\right]_+\nonumber\\
&+& {1+z^2\over 1-z}
\left (
8 \Li_3(-z)- 4 \Li_2(-z)\ln(z)+{9\over 2}\ln(1-z)\ln^2(z)
+{\pi^2 \over 6} \ln(z)
\right )
+ {2(1+7z^2)\over 1-z}\Li_3(z) -{4\zeta(3)\over 1-z}
\nonumber\\
&-& {7-z^2\over 1-z} \left ( \Li_3(1-z)+\Li_2(z)\ln(1-z) \right )
- {8z^2\over 1-z}\Li_2(z)\ln(z)
-{9z^2+7\over 12(1-z)}\ln^3(z)
-{17z^2+25\over 2(1-z)}\ln^2(1-z)\ln(z)
\nonumber\\
&-&4(z+1)\ln^3(1-z)
- {8z(z+1)\over 1-z}\Li_2(-z)
-{21z^2+22z-22\over 3(1-z)}\Li_2(z)
+
{33z^4-40z^3+3z^2-18z+6\over 12(1-z)^3}\ln^2(z)\nonumber\\
 &-& {21z^2+46z-13\over 3(1-z)}\ln(1-z) \ln(z)
-{8z(z+1)\over 1-z}\ln(z+1) \ln(z) -2(z+6)\ln^2(1-z)
 -{3z^2-5\over 2(1-z)}\pi^2 \ln(1-z)
\nonumber\\
&-&{63z^4-144z^3+40z^2-32z+9\over 6(1-z)^2(z+1)}\ln(z)
+ {16z-5\over 2}\ln(1-z)
+{3z^2-10z+6\over 6(1-z)}\pi^2+{21z^2-26z+45\over 6(1-z)} ,
\end{eqnarray}
\end{widetext}
\vskip 1mm
\begin{widetext}
\begin{eqnarray}
&& F^{(C_AC_F)}_ {\Q} =
\left \{
{11\over 8}\delta(1-z)
+ {11\over 6}\left[{1\over 1-z}\right]_+
 -{11\over 12}(1+z)\right\}L^2 +\left\{ \delta(1-z) \left({35\over 8}
+{11\pi^2\over 18}-3\zeta(3)\right)
\right. \nonumber \\
&& \left. -{22\over 3}\left[{\ln(1-z)\over 1-z}\right]_+
+\left({34\over 9}-{\pi^2\over 3}\right)\left[{1\over 1-z}\right]_+
+ {1+z^2\over 2(1-z)}\ln^2(z)+{17+5z^2 \over 6(1-z)}\ln(z)
+{11\over 3}(1+z)\ln(1-z)
\right. \nonumber \\
&& \left. +{\pi^2\over 6}(1+z)
+{43 -77z \over 9}\right \} L
+ \delta(1-z)\left({1141\over 288} +{41\over 54}\pi^2
+\pi^2\ln(2)+{\pi^4\over 36}-{\zeta(3)\over 2}\right)
+{22\over 3}\left[{\ln^2(1-z)\over 1-z}\right]_+
\nonumber\\
&-&\left({68\over 9}-{2\pi^2\over 3}\right)\left[{\ln(1-z)\over
1-z}\right]_+ +\left({55\over 27}+{\pi^2\over 3}
-9\zeta(3)\right)\left[{1\over 1-z}\right]_+ + {1+z^2\over 1-z}
\left (-5 \Li_3(z) -4\Li_3(-z) -\Li_3(1-z)
\right. \nonumber \\
&& \left.
+{\ln^3(z)\over 4}
+ \Li_2(z) ( 2\ln(z)+\ln(1-z) )
+   2  \Li_2(-z)\ln(z)
-\ln(1-z) \ln^2(z)+ \ln^2(1-z) \ln(z) +{\pi^2\over 6}\ln(z)
\right )
\nonumber \\
&& -{5z^2-13\over 2(1-z)}\zeta(3)+{\pi^2(z^2-3)\over 6(1-z)}\ln(1-z)
+ {33z^2-2z+8\over 6(1-z)} \Li_2(z)
+\frac{4z(1+z)}{1-z}\Li_2(-z)
-{ 2z^2+4z+7\over 3(1-z)}\ln(1-z)\ln(z)
\nonumber \\
&&
-{13z^4-18z^3+2z^2-2z-11\over 24(1-z)^3}\ln^2(z)
+  {4z(z+1)\over 1-z}\ln(z+1)\ln(z)
-{(41z+47)\over 12}\ln^2(1-z)-{6z^2-4z+3\over 12(1-z)}\pi^2
\nonumber\\
&+&{173z-37\over 18}\ln(1-z)
+{43z^4-159z^3+96z^2-105z+29\over 18(1-z)^2(z+1)}\ln(z)
 +{2z^2-39z-53\over 27(1-z)} ,
\end{eqnarray}
\end{widetext}

\vskip 2mm

\begin{widetext}
\begin{eqnarray}
&& F^{(C_FT_R)}_ {\Q} =
\left\{
-{1\over 2}\delta(1-z)
-{2\over 3}\left[{1\over 1-z}\right]_+
+(1+z)\ln(z)-{4z^2+z-5\over 6}
+{2\over 3z}\right \}L^2\nonumber\\
&+&\left\{
-\left({3\over 2}+{2\pi^2 \over 9}\right)\delta(1-z)
+{8\over 3}\left[{\ln(1-z)\over 1-z}\right]_+
-{8\over 9}\left[{1 \over 1-z}\right]_+
+ (1+z)\ln^2(z) + {(8z^3+17z^2-12z-17) \over 3(1-z)}\ln(z)
\right.
\nonumber\\
&-& \left.
{4\over 3}(z+1)\ln(1-z)
+{ 56z^2+52z-80 \over 9}-{20\over 9z}\right \}L
+\delta(1-z) \left({3139\over 648}-{\pi^2\over 3}+{2\over 3}\zeta(3)
\right) +{56\over 27}\left[{1\over 1-z}\right]_+
\nonumber \\
&+&(z+1)\left( -8\Li_3(z) -16\Li_3(-z)+4\ln(z)(2 \Li_2(-z) +\Li_2(z))
 +{\ln^3(z) \over 6}
-4\zeta(3) \right)\nonumber\\
&+&{4(z-1)(z^2+4z+1)\over 3z}\left(2\Li_2(-z) + \Li_2(z)
+\ln(1-z)\ln(z) +2\ln(z+1)\ln(z) \right)
+{8z^3+17z^2-12z-17\over 12(1-z)}\ln^2(z)\nonumber\\
&+&{20z^9+45z^7+40z^8-339z^6-215z^5-345z^2+921z^4
-457z^3+159z+43\over 9(z+1)^3(1-z)^4}\ln(z)\nonumber\\
&+&{400z^7+93z^6   -1970z^5+963z^4+1396z^3
-765z^2-1170z+669 \over 54(z+1)^2(1-z)^3} + {56\over 27z} ,
\end{eqnarray}
\end{widetext}

\vskip 2mm

\begin{widetext}
\begin{eqnarray}
&& F^{(C_FT_Rn_l)}_ {\Q} =
\left\{
-{1\over 2}\delta(1-z)
-{2\over 3}\left[{1\over 1-z}\right]_+
+{1+z\over 3}\right \} L^2
+\left\{
- \left ( {3 \over 2} + {2\pi^2 \over 9}\right)\delta(1-z)
+ {8\over 3}\left[{\ln(1-z)\over 1-z}\right]_+
\right. \nonumber \\
&-& \left.
{8\over 9} \left[{1\over 1-z}\right]_+
- {2 (1+z^2) \over 3(1-z)}\ln(z)
-{4\over 3}(1+z)\ln(1-z)+{16z -8\over 9}\right \}L
-\delta(1-z) \left({173\over 72}+{8\over 27}\pi^2+2\zeta(3)\right)
\nonumber\\
&&
-{8\over 3}\left[{\ln^2(1-z)\over 1-z}\right]_+
+{16\over 9}\left[{\ln(1-z)\over 1-z}\right]_+
+{4\over 27}\left[{1\over 1-z}\right]_+
+ {1+z^2\over 1-z} \left (
{4\over 3}\ln(1-z)\ln(z) - {\ln^2(z)\over 6} \right )
\nonumber\\
&+& {4\over 3}(z+1)\ln^2(1-z) - {5z^2-12z+5\over 9(1-z)}\ln(z)
-{16\over 9}(2z-1)\ln(1-z)+{19z-23\over 27} .
\end{eqnarray}
\end{widetext}
\vskip 1mm

The results for the functions $d^{(2)}_{\Q}$ and
$d^{(2)}_{\overline{\Q}}$ presented above satisfy the fermion
number conservation condition Eq.(\ref{FNC}). The two
functions  are not separately integrable over the interval
$0\leq z\leq 1$ because of the  terms $\sim 1/z$;  however, all
such terms cancel in the difference.  The integral of the function
$F^{(C_FT_Rn_l)}_ {\Q}$ over  $0\leq z\leq 1$
vanishes as expected, since $d^{(2)}_{\overline{\Q}}$ has no
terms proportional to $n_l$.

Our result can be compared with the large-$\beta_0$ approximation
of Ref. \cite{CacciariGardi}.  We have checked that the
${\cal{O}}(\beta_0 \as^2)$ term in that paper  coincides with the
$n_l$-dependent part of our result Eq.(\ref{d2Qresult}). Another
check of our result is provided by the soft, $z \to 1$, limit. In
such kinematic regime, almost all the energy of the initial parton
is transferred to the observed heavy quark in the final state; the
energy radiated away in the fragmentation process is small. This
leads to large contributions to $\D$, of the form $\alpha_s^n
\ln^{k}(1-z)/(1-z)$ with $k\leq 2n-1$, which can be resummed using
the soft gluon resummation formalism  \cite{CC}. When the resummed
result in \cite{CC} is expanded in powers of $\as$ through ${\cal
O}(\alpha_s^2)$, an approximation for $d^{(2)}$, valid for  $z
\approx 1$, is obtained.  We can compare those predictions with
the explicit calculation presented in this paper. It is convenient
to perform such comparison considering the Mellin moments of
$d^{(2)}(z)$. Recall that the  Mellin transform  is defined as:
\begin{equation}
d_{\Q}^{(2)}(N) = \int_0^1 dz~z^{N-1}d_{\Q}^{(2)}(z).
\end{equation}
A useful summary of Mellin moments is given in
Ref.\cite{Blumlein}. The soft, $z \to 1$ limit corresponds to
$N \to \infty$ limit in the Mellin space.  In what follows,
we present the  non-vanishing
asymptotics  of $d_{\Q}^{(2)}(N)$,  for $N \to \infty$:
\begin{widetext}
\begin{eqnarray}
&& F^{(C_F^2)}_ {\Q} \to 2\ln^4(N)+(4L+8\gamma_{\rm
E}-4)\ln^3(N)+\left(2L^2+(12\gamma_{\rm E}-7)L
+{2\over 3}\pi^2-2+12\gamma_{\rm E}^2-12\gamma_{\rm E}\right)\ln^2(N)
\nonumber\\
&+&\left( (4\gamma_{\rm E}-3)L^2+\left({2\over 3}\pi^2-1+12\gamma_{\rm E}^2
-14\gamma_{\rm E}\right)L +{2\over 3}(\pi^2-6+6\gamma_{\rm E}^2
-6\gamma_{\rm E})(2\gamma_{\rm E}-1)\right)\ln(N) \nonumber\\
&+& {1\over 8}(4\gamma_{\rm E}-3)^2L^2+\left(-\pi^2+{27\over
8}+6\zeta(3) -\gamma_{\rm E}-7\gamma_{\rm E}^2+ {2\gamma_{\rm
E}\pi^2 \over 3}+4\gamma_{\rm E}^3\right)L +4\gamma_{\rm
E}-2\pi^2\ln(2)-4\gamma_{\rm E}^3
\nonumber\\
&-&{3\over 2}\zeta(3)+{1\over 4}\pi^2+2\gamma_{\rm E}^4+{2\over 3}
\pi^2\gamma_{\rm E}^2-{2\over 3}\pi^2\gamma_{\rm E}-2\gamma_{\rm E}^2-
{11\over 180}\pi^4+{241\over 32} , \nonumber\\
&& F^{(C_AC_F)}_ {\Q} \to -{22\over 9}\ln^3(N)
+\left(-{11\over 3}L-{22\over 3}\gamma_{\rm E}-{34\over 9}
+{\pi^2\over 3}\right)\ln^2(N)
+\left(-{11\over 6}L^2+\left(-{34\over 9}+{\pi^2\over 3}
-{22\gamma_{\rm E}\over 3} \right)L
\right. \nonumber \\
&&\left.
 -{55\over 27}-{14\over 9}\pi^2
+9\zeta(3)-{68\over 9}\gamma_{\rm E}+{2\over 3}\pi^2\gamma_{\rm E}
-{22\over 3}\gamma_{\rm E}^2\right)\ln(N)
+ \left(-{11\gamma_{\rm E} \over 6}+{11\over 8}\right)L^2
+ \left( -{11\gamma_{\rm E}^2\over 3}+{35\over 8}-3\zeta(3)
\right. \nonumber \\
&&\left. -{34\gamma_{\rm E}\over 9}+{\gamma_{\rm E}\pi^2\over
3}\right)L - {22\gamma_{\rm E}^3\over 9}+\pi^2\ln(2) +{1141\over
288}-{34\gamma_{\rm E}^2\over 9}+{7\pi^2\over
54}+{\pi^2\gamma_{\rm E}^2\over 3}+{\pi^4\over 12}-{14
\pi^2\gamma_{\rm E}\over 9}-{97\zeta(3)\over 18}-{55\gamma_{\rm
E}\over 27}+9\zeta(3)\gamma_{\rm E} ,  \nonumber\\
&& F^{(C_FT_R)}_ {\Q} \to {4\over 3}\ln^2(N) L +\left( {2\over
3}L^2+{8\over 9}(3\gamma_{\rm E}+1)L-{56\over 27}\right)
\ln(N)+\left({2\gamma_{\rm E}\over 3}-{1\over 2}\right)L^2 +
\left({4\gamma_{\rm E}^2\over 3}+ {8\gamma_{\rm E}\over 9}-{3\over
2}\right)L
\nonumber\\
&&-{56\gamma_{\rm E}\over 27}+{3139\over 648}-
{\pi^2\over 3}+{2\over 3}\zeta(3) ,\nonumber\\
&& F^{(C_FT_Rn_l)}_ {\Q} \to {8\over 9}\ln^3(N)+
\left({4\over 3}L+{8\over 9}+{8\over 3}\gamma_{\rm E}\right)\ln^2(N)
+\left({2\over 3}L^2+\left({8\gamma_{\rm E}\over 3}+{8\over 9}
\right)L+{16\over 9}\gamma_{\rm E}-{4\over 27}+{8\over 3}
\gamma_{\rm E}^2+{4\over 9}\pi^2\right)\ln(N)\nonumber\\
&&+\left({2\gamma_{\rm E}\over 3}-{1\over 2}\right)L^2
+\left({4\gamma_{\rm E}^2\over 3}- {3\over 2}+{8\gamma_{\rm
E}\over 9}\right)L +{8\over 9}\gamma_{\rm E}^3+{8\over
9}\gamma_{\rm E}^2+\left(-{4\over 27}+ {4\over
9}\pi^2\right)\gamma_{\rm E}-{173\over 72}- {4\over
27}\pi^2-{2\over 9}\zeta(3).
\end{eqnarray}
\end{widetext}
Here $\gamma_{\rm E}$ is the Euler constant.

Comparing these results with Ref.\cite{CC}, we find full agreement
for all terms $\as^2(\mu_0) \ln^k(N), k\ge 2$ that were
investigated there, provided that we use the matching relation for
the coupling constants evolving with $n_l$ and $n_{l}+1$ flavors.
The results for the subleading $\ln^k(N),~k=0,1$ terms are new;
they can be used to extract the coefficient $H^{(2)}$ (cf. Eq.(70)
in \cite{CC}) needed to extend the soft gluon resummation  for
$D^{\rm ini}$ to next-to-next-to-leading logarithmic accuracy.

\section{Conclusions}

In this paper, we derived the initial condition for the heavy
quark perturbative fragmentation function through  ${\cal
O}(\alpha_s^2)$. The importance of this result is twofold. First,
it enables us  to derive energy distributions of heavy quarks
produced in hard scattering processes from pure massless
calculations. In addition, it can be used to resum large collinear
logarithms through NNLL level, using the DGLAP evolution equation.

To derive the initial condition, we make use of the
process-independent approach that was recently proposed in
\cite{CC,KL}. We suggest a simple modification of the original
proposal that renders the methods for multi-loop calculations
applicable to the calculation of the initial condition of the
perturbative fragmentation function.

In the context of QED, an interesting potential application of the
initial condition, derived in this paper, is the calculation of
the electron energy spectrum in muon decay. Currently, this
observable is being very accurately measured in TWIST experiment.
The goal of the experiment is to extract Michel parameters
\cite{michel} from the electron energy spectrum with a relative
precision at the level of $10^{-4}$. Known results on ${\cal
O}(\alpha^2 \ln^n(m_\mu/m_e)),~n=1,2$ terms \cite{mudec,technique}
suggest that ${\cal O}(\alpha^2)$ corrections without logarithmic
enhancement might be important for an unambiguous interpretation
of the experimental result. With the initial condition for the
fragmentation function at hand, one can obtain the electron energy
spectrum in muon decay by  computing the energy spectrum for
massless electron, removing collinear divergencies by ${\overline
{\rm MS}}$ renormalization and convoluting  decay rate, obtained
in this way, with the initial condition for the fragmentation
function derived in this paper. This approach simplifies the
calculation of the electron energy spectrum in muon decay. Similar
methods can be used to extend the results of Ref.\cite{Blum} on
QED corrections to deep-inelastic scattering.

In the context of collider physics, there are many applications
for the initial condition of the perturbative heavy quark
fragmentation function. First, the $B$-meson energy spectrum was
measured by ALEPH, OPAL and SLD collaborations; this is currently
the primary source of information about the $b$-quark
fragmentation function. It has been observed \cite{carlo} that
inclusion of higher order QCD corrections reduces the significance
of non-perturbative effects in the bottom quark fragmentation. The
knowledge of $\D$ through ${\cal O}(\alpha_s^2)$, permits a
reanalysis of $B$-meson fragmentation with  the NNLL accuracy.
Another example of potential application is an accurate prediction
of the $b$-quark energy spectrum in top quark decay. This is
currently known to ${\cal O}(\alpha_s)$ \cite{Corcella:2001hz}.
The calculation presented in this paper offers the possibility to
extend this analysis to ${\cal O}(\alpha_s^2)$.

{\bf Acknowledgments.}
This research is supported by the DOE under
contract DE-FG03-94ER-40833 and the Outstanding Junior
Investigator Award DE-FG03-94ER40833, by the National Science
Foundation under Grant No. PHY99-07949 and by the start up funds
of the University of Hawaii. We are grateful to the KITP, UC Santa
Barbara, for its hospitality during the completion of this work.

\end{document}